# Spatial Autocorrelation and Verdoorn Law in the Portuguese NUTs III

**Vítor João Pereira Domingues Martinho**

Escola Superior Agrária, Instituto Politécnico de Viseu, Quinta da Alagoa,

Estrada de Nelas, Ranhados, 3500 - 606 VISEU

Centro de Estudos em Educação, Tecnologias e Saúde (CI&DETS)

e-mail: vdmartinho@esav.ipv.pt

# Spatial Autocorrelation and Verdoorn Law in the Portuguese NUTs III

**Abstract**

This study analyses, through cross-section estimation methods, the influence of spatial effects in productivity (product per worker), at economic sectors level of the NUTs III of mainland Portugal, from 1995 to 1999 and from 2000 to 2005 (taking in count the data availability and the Portuguese and European context), considering the Verdoorn relationship. From the analyses of the data, by using Moran I statistics, it is stated that productivity is subject to a positive spatial autocorrelation (productivity of each of the regions develops in a similar manner to each of the neighbouring regions), above all in services. The total sectors of all regional economy present, also, indicators of being subject to positive autocorrelation in productivity. Bearing in mind the results of estimations, it can been that the effects of spatial spillovers, spatial lags (measuring spatial autocorrelation through the spatially lagged dependent variable) and spatial error (measuring spatial autocorrelation through the spatially lagged error terms), influence the Verdoorn relationship when it is applied to the economic sectors of Portuguese regions. The results obtained for the two periods are different, as expected, and are better in second period, because, essentially, the European and national public supports (Martinho, 2011).

## 1. Introduction

This study seeks to test Verdoorn's Law (using product per worker as a proxy for productivity) for each of the economic sectors of regions (NUTs III) of mainland Portugal from 1995 to 1999 and from 2000 to 2005, through techniques of cross-section spatial econometrics. The mainly conclusion is that productivity is subject to a positive spatial autocorrelation in the economic sectors of the Portuguese regions, with some differences in each particular situations. On the other hand the effects of spatial spillovers, spatial lags and spatial error, influence the Verdoorn relationship when it is applied to this economic sectors. In the second period the results are different, as is expected, because the context in Portugal is distinct and in our point of view the results are better. Portugal benefited from a European and national public support to modernize the economic sectors, since the Portuguese entrance, in 1986, to the, at that time, called European Economic Community. In the first considered period, in this work (1995 to 1999), Portugal benefited from the second Community Support Framework and in the second period (2000 to 2005) benefited from the third Community Support Framework negotiated under the Agenda 2000. This third Community Support Framework was very important to Portugal because appear in a phase where the national public services are more developed and efficient, the access to the information is easier, because the internet, and in the other hand the structural programs were better defined. In this way the national Portuguese economy benefited from this third Community Support Framework in the economic sectors modernization and the results in this work show this.

In a point of view technical and public policy decisions these are important results, because different reasons. The existence of spatial autocorrelation means that the policies must be directed to a strategic sector in a strategic place, because from here the effects spread to the other sectors in the other places. When the public resources scarce and they are not enough for everyone and for everywhere this must be take in count. The fact of the spatial autocorrelation affect positively the verdoorn law is a good new, because we know this law captures positive effects in the economic growth, so in this way we have good spread effects for the economy. The better results for the second period signify that Portugal needs the national public and European supports to modernize the Portuguese economy. Portugal has significant structural problems which must be solved as soon as possible, so we can compete in globalized world.

The consideration of spatial effects at a regional level is becoming increasingly frequent and the work of Anselin (1988), among others, has contributed to this. Anselin (2002) has reconsidered a number of conceptual matters related to implementing an explicit spatial perspective in applied econometrics. The influence of neighbouring locations (parishes, councils, districts, regions, etc) in the development of a particular area, through the effects of spatial spillovers, is increasingly considered in more recent empirical studies, a fact which has been highlighted by Anselin (2003a and 2007). The several works of Luc Anselin refer to the inclusion of spatial effects as being important from an econometric point of view. If the underlying data arises from processes which include a spatial dimension, and this is omitted, the estimators are either biased and inconsistent or inefficient depending on whether the error or the lag model is the underlying data generating process.

Following on from these studies, the development of productivity of a particular region, for example, can be influenced by the development of productivity in neighbouring regions, through external spatial factors. The existence or non-existence of these effects can be determined through a number of techniques which have been developed for spatial econometrics, where Anselin, among others, in a number of studies has made a large contribution. Paelinck (2000) has brought a number of theoretical contributions to the aggregation of models in spatial econometrics, specifically concerning the structure of parameters. Anselin and Moreno (2003) considered a group of specification tests based on the method of Maximum Likelihood to test the questions related with the spatial error component. Anselin (2003b) has presented a classification of specification for models of spatial econometrics which incorporates external spatial factors. Baltagi et al. (2003) has sought to present improvements in specification tests (testing whether the more correct specification of models is with the spatial lag component or the spatial error component) LM (Lagrange Multiplier), so as to make it more adaptable to spatial econometrics. Anselin et al. (1996) has proposed a simple, robust diagnostic test, based on the OLS method, for the spatial autocorrelation of errors in the presence of spatially lagged dependent variables and vice-versa, applying the modified LM test developed before. This test was, also, after proposed by Florax et al. (2006).

Few studies have been carried out on analysing the conditional productivity convergence with spatial effects and none, at least to our knowledge, concerning productivity being dispersed by the various economic sectors. Fingleton (2001), for example, has found a spatial correlation in productivity when, using the data from 178 regions of the European Union, he introduced spillover effects in a model of endogenous growth. Abreu et al. (2004) have investigated the spatial distribution of growth rates in total factor productivity, using exploratory analyses of spatial data and other techniques of spatial econometrics. The sample consists of 73 countries and covers the period 1960-2000.

On the other hand, there is some variation in studies analysing conditional convergence of product with spatial effects. Ramajo et al. (2008), for example, have examined the absolute and conditional convergence hypothesis across European regions from the period 1981 to 1996. Costantini et al. (2006) examines stochastic convergence in real per capita GDP for Italian regions using recent non-stationary panel data methodologies over the period 1951 to 2002. Lundberg (2006) has tested the hypothesis of conditional convergence with spatial effects between 1981 and 1990 and, in contrast to previous results, found no clear evidence favouring the hypothesis of conditional convergence. On the contrary, the results foresaw conditional divergence across municipalities located in the region of Stockholm throughout the period and for municipalities outside of the Stockholm region during the 1990s.

Spatial econometric techniques have also been applied to other areas besides those previously focused on. Longhi et al. (2006), for example, have analysed the role of spatial effects in estimating the function of salaries in 327 regions of Western Germany during the period of 1990-1997. Anselin et al. (2004) have analysed the economic importance of the use of analyses with spatial regressions in agriculture in Argentina. Kim et al. (2003) have measured the effect of the quality of air on the economy, through spatial effects, using the metropolitan area of Seoul as a case study. Chen et al (2010) use an econometric model that takes into account the spatial dependence, and allow the effect of access to differ for a person depending on whether he or she lives in a low-income community or peer group.

In this way there are several works taking in count the referred aspects, as we will see in the next section for several countries. This work has as principal contributions the analysis of these questions to the Portuguese reality and is the only study we know, in this perspective, for the economy of the Portugal mainland.

To do so, the rest of the study is structured as follows: in section 2 presents the methodology. Section 3 presents the data sources and the data description while section 4 does the empirical results analysis. Finally, section 5 provides a summary of the results and discusses their policy implications.

## 2. Methodology considerations

In this section we present the following: some previous theoretical considerations, namely about Verdoorn Law and Spatial Econometric, the model, the data and some explanations about the software used in this work.

### 2.1. Some theoretical considerations

Verdoorn detected that there was an important positive relationship between the growth of productivity of work and the growth of output. He defended that causality goes from output to productivity, with an elasticity of approximately 0.45 on average (in cross-section analyses), thus assuming that the productivity of work is endogenous.

After, in the sixties, Kaldor redefined this Law and its intention of explaining the causes of the poor growth rate in the United Kingdom, defended that there was a strong positive relationship between the growth of work productivity (p) and output (q), so that, p=f(q). Or alternatively, between the growth of employment (e) and the growth of output, so that, e=f(q).

There have been various studies carried out concerning Verdoorn's Law considering the possibility of there being spatial spillover effects.

Bernat (1996), for example, when testing Kaldor's three laws of growth in regions of the USA from the period of 1977 to 1990, distinguished two forms of spatial autocorrelation: spatial lag and spatial error. Spatial lag is represented as follows: $y = \rho W y + X\beta + \varepsilon$, where y is the vector of endogenous variable observations, , W is the distance matrix, X is the matrix of endogenous variable observations, $\beta$ is the vector of coefficients, $\rho$ is the self-regressive spatial coefficient and $\varepsilon$ is the vector of errors. The coefficient $\rho$ is a measurement which explains how neighbouring observations affect the dependent variable. The spatial error model is expressed in the following way: $y = X\beta + \mu$, where spatial dependency is considered in the error term $\mu = \lambda W \mu + \xi$.

To resolve problems of spatial autocorrelation, Fingleton and McCombie (1998) considered a spatial variable which would capture the spillovers across regions, or, in other words, which would determine the effects on productivity in a determined region i, on productivity in other surrounding regions j, as the distance between i and j.

Fingleton (1999), has developed an alternative model and Fingleton (2007) analyse this question in a point of view more structured and more completed. A potential source of errors of specification in spatial econometric models comes from spatial heterogeneity (Lundberg, 2006). There are typically two aspects related to spatial heterogeneity, structural instability and heteroskedasticity. To prevent these types of errors of specification and to test for the existence of spatial lag and spatial error components in models, the results are generally complemented with specification tests. One of the tests is the Jarque-Bera test which tests the stability of parameters. The Breuch-Pagan and Koenker-Bassett, in turn, tests for heteroskedasticity. The second test is the most suitable when normality is rejected by the Jarque-Bera test. To find out if there are spatial lag and spatial error components in the models, two robust Lagrange Multiplier tests are used ($LM_E$ for "spatial error" and $LM_L$ for "spatial lag"). In brief, the $LM_E$ tests the null hypothesis of spatial non-correlation against the alternative of the spatial error model ("lag") and $LM_L$ tests the null hypothesis of spatial non-correlation against the alternative of the spatial lag model to be the correct specification.

According to the recommendations of Florax et al. (2006) and using the so-called strategy of classic specification, the procedure for estimating spatial effects should be carried out in six steps: 1) Estimate the initial model using the procedures using OLS; 2) Test the hypothesis of spatial non-dependency due to the omission spatially lagged variables or spatially autoregressive errors, using the robust tests $LM_E$ and $LM_L$; 3) If none of these tests has statistical significance, opt for the estimated OLS model, otherwise proceed to the next step, 4) If both tests are significant, opt for spatial lag or spatial error specifications, whose test has greater significance, otherwise go to step 5;; 5) If $LM_L$ is significant while $LM_E$ is not, use the spatial lag specification; 6) If $LM_E$ is significant while $LM_L$ is not, use the spatial error specification.

A test usually used to indicate the possibility of global spatial autocorrelation is the Moran's I test. On the other hand Moran's I local autocorrelation test investigates if the values coming from the global autocorrelation test are significant or not.

**2.2. The model**

Bearing in mind the previous theoretical considerations, what is presented next is the model used to analyse Verdoorn's law with spatial effects, at a regional and sector level.

As a result, to analyse Verdoorn's Law in the regional economic sectors the following model can be used:

$$p_{it} = \rho W_{ij} p_{it} + \gamma q_{it} + \varepsilon_{it}, \text{ Verdoorn's equation with spatial effects} \qquad (1)$$

where p are the rates of growth of sector productivity across various regions, W is the matrix of distances, q is the rate of growth of output, , $\gamma$ is Verdoorn's coefficient which measures economies to scale (which it is hoped of values between 0 and1), $\rho$ is the autoregressive spatial coefficient (of the spatial lag component) and $\varepsilon$ is the error term (of the spatial error component, with, $\varepsilon = \lambda W \varepsilon + \xi$ ). The indices i, j and t, represent the regions being studied, the neighbouring regions and the period of time respectively.

**2.3. The data**

We obtained data about the output and the employment for the economic sectors of the Portuguese regions (NUTS III). The data were obtained in the regional accounts of the National Statistics Institute. This data are available online at the following internet site: http://www.ine.pt/xportal/xmain?xpid=INE&xpgid=ine_publicacoes.We consider the two periods (from 1995 to 1999 and from 2000 to 2005) because the Agenda 2000 which is an action program whose main objectives are to strengthen Community policies and to give the European Union a new financial framework for the period 2000-2006 with a view to enlargement. We think at this fine regional level these questions have much importance.

**2.4. The software**

Some preliminary considerations about the informatics programme used to the data analysis and for estimations. We used the GeoDa, that is a recent computer programme with an interactive environment that combines maps with statistical tables, using dynamic technology related to Windows (Anselin, 2006). In general terms, functionality can be classified in six categories: 1) Manipulation of spatial data; 2) Transformation of data; 3) Manipulation of maps; 4) Construction of statistical tables; 5) Analysis of spatial autocorrelation; 6) Performing spatial regressions. All instructions for using GeoDa are presented in the following internet site http://geodacenter.asu.edu/.

## 3. Exploratory data analysis

The analysis sought to identify the existence of Verdoorn's relationship by using Scatterplots and spatial autocorrelation with the Moran Scatterplots for global spatial autocorrelation and LISA Maps for local spatial autocorrelation. In this analysis of data and the estimations which will be carried out in part four of this study, the dependent variable of the equation used to test Verdoorn's Law is presented in average growth rates for the period considered for cross-section analysis.

### 3.1. Scatterplots

With the eight (Figure 1 and 2) Scatterplots presented below we pretend to make an analysis of the existence of a correlation between growth of productivity and product growth under Verdoorn's Law (equation (1)), for each of the economic sectors (agriculture, industry, services and the total of all sectors) of Portuguese NUTs III (28 regions), with average values for the period 1995 to 1999 and from 2000 to 2005. Is expected that the Industry to be the sector with stronger relationship between this two variables.

To analyse the Scatterplots we confirm what is defended by Kaldor, or, in other words, Verdoorn's relationship is stronger in industry (a sign of being the sector with the greatest scaled income, although the underlying value is far too high) and weaker in other economic sectors (an indication that these sectors have less scaled income). On the other, the agriculture is an exception here (since there is evidence of quite high scaled income, which is contrary to what was expected when considering the theory), due to the restructuring which it has undergone since Portugal joined the EEC, with the consequent decrease in population active in this sector which is reflected in increased productivity.

**Figure 1: "Scatterplots" of Verdoorn's relationship for each of the economic sector (cross-section analysis, 28 regions, 1995-1999)**

a) Agriculture                                           b) Industry

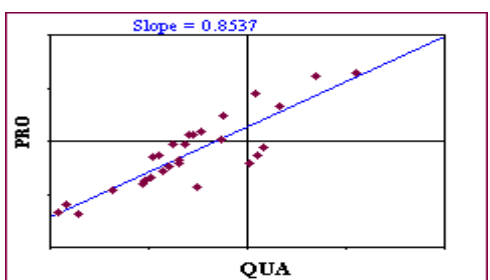

c) Services                                         d) All sectors

Note: PRO = Productivity;

QUA = Product.

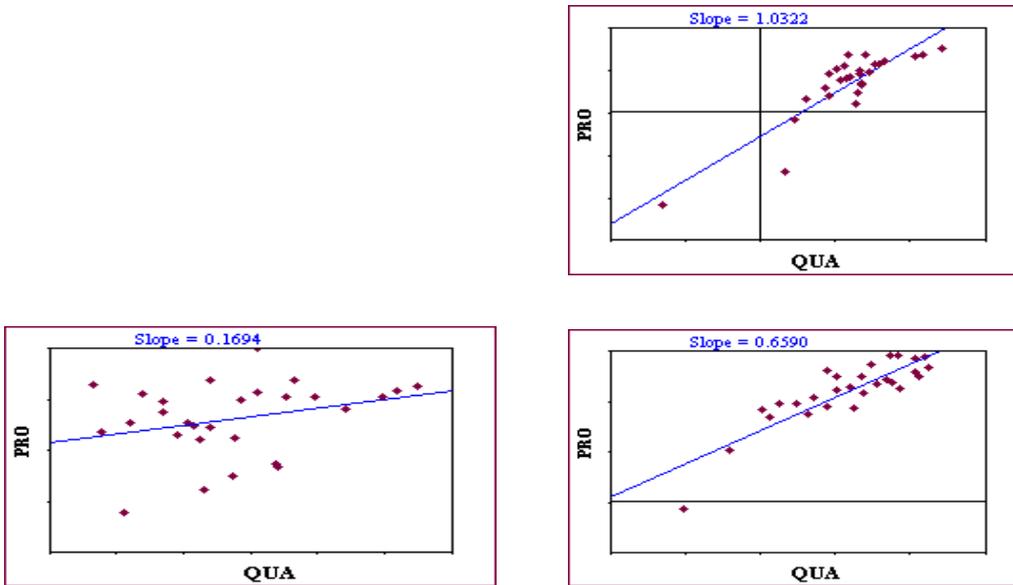

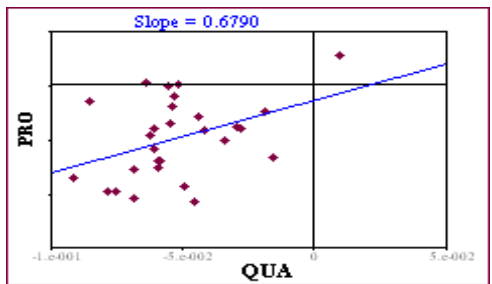
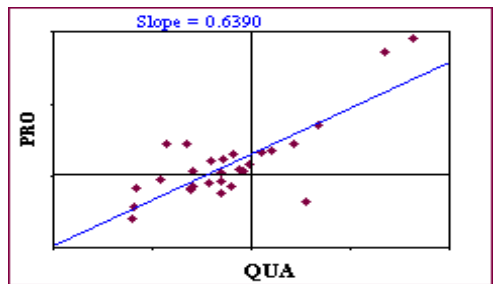

Figure 2: "Scatterplots" of Verdoorn's relationship for each of the economic sector (cross-section analysis, 28 regions, 2000-2005)

a) Agriculture                            b) Industry

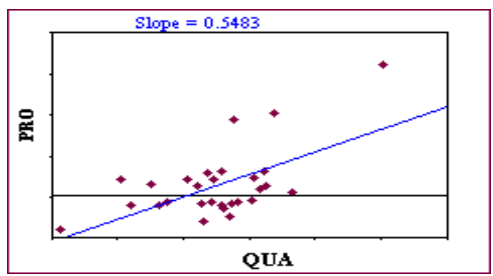
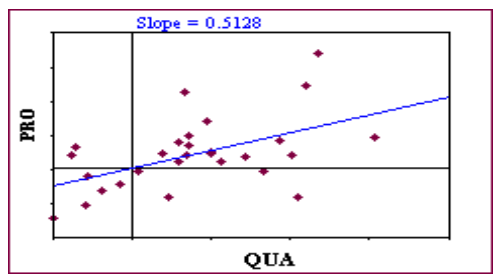

c) Services                                        d) All sectors

Note: PRO = Productivity;

QUA = Product.

## 3.2. Moran Scatterplots

The eight (Figure 3 and 4) Moran Scatterplots which are presented below concerning the dependent variable (average growth rates of productivity in the period 1995 to 1999 and from 2000 to 2005), constructed by the equation of Verdoorn's Law, show Moran's I statistical values for each of the economic sectors and for the totality of sectors in the 28 NUTs in mainland Portugal. We pretend to explore the existence of spatial autocorrelation with the Moran´s I statistics. There are spatial spillovers when the Moran´s I present positive values. The matrix $W_{ij}$ used is the matrix of the distances between the regions up to a maximum limit of 97 Km. This distance appeared to be the most appropriate to the reality of Portuguese NUTs III, given the diverse values of Moran's I obtained after various attempts with different maximum distances. For example, for services which, as we shall see, is the sector where the Moran's I has a positive value (a sign of spatial autocorrelation), this value becomes negative when the distances are significantly higher than 97 Km, which is a sign that spatial autocorrelation is no longer present. On the other hand, the connectivity of the distance matrix is weaker for distances over 97 Km. Whatever the case, the choice of the best limiting distance to construct these matrices is always complex.

Would be good if we had more observations, but is difficult to find to a finer spatial unity. Anyway the results obtained are consistent with the Portuguese reality taking into account another works about regional growth, because the spataila analysis are not so dependent of the number of obervations This has some practical and theoretical explanations, but the principal is about in some cases the heterogeneity of the spatial unities. In this way, more spatial unities could be a signal of more diversity.

An analysis of the Moran Scatterplots demonstrates that it is principally in services that a global spatial autocorrelation can be identified and that there are few indicators that this is present in the totality of sectors, since Moran's I value is positive.

**Figure 3: "Moran Scatterplots" of productivity for each of the economic sectors (cross-section analysis, 28 regions, 1995-1999)**

a) Agriculture          b) Industry

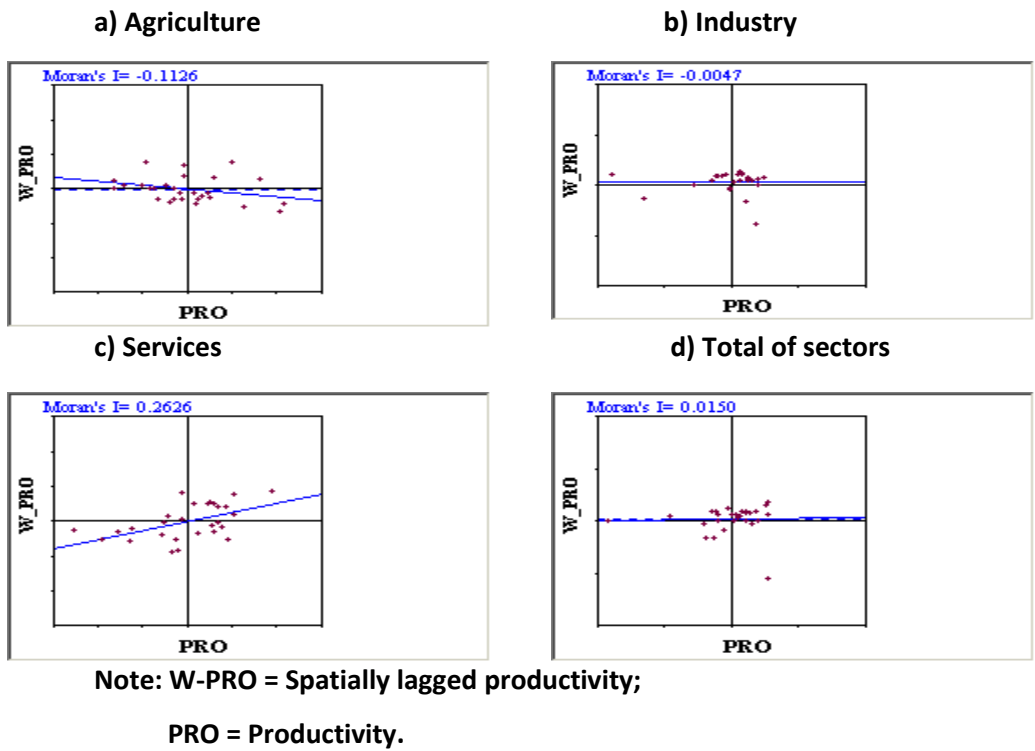

c) Services          d) Total of sectors

Note: W-PRO = Spatially lagged productivity;

PRO = Productivity.

Figure 4: "Moran Scatterplots" of productivity for each of the economic sectors (cross-section analysis, 28 regions, 2000-2005)

a) Agriculture          b) Industry

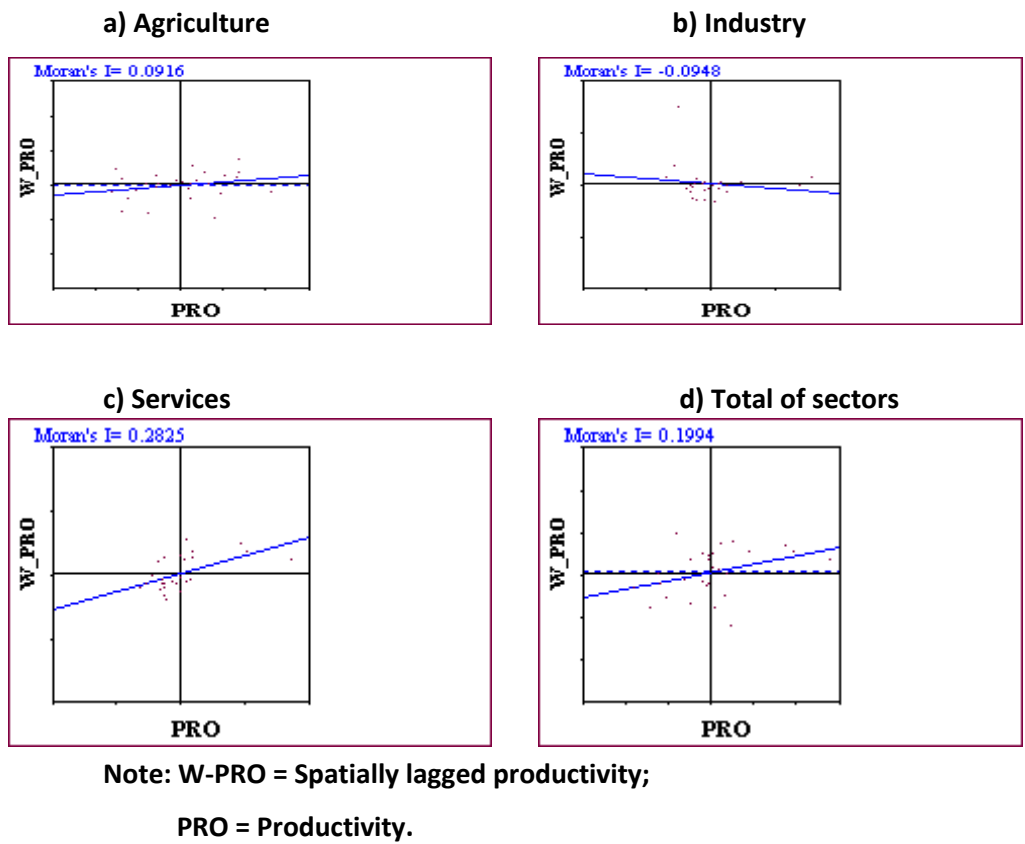

c) Services          d) Total of sectors

Note: W-PRO = Spatially lagged productivity;

PRO = Productivity.

3.3. LISA maps

Below we pretend to make an analysis of the existence of local spatial autocorrelation with eight LISA Maps (Figure 5 and 6), investigated under spatial autocorrelation and its significance locally (by NUTs III). The NUTs III with "high-high" and "low-low" values, correspond to the regions with positive spatial autocorrelation and with statistical significance, or, in other words, these are cluster regions where the high values ("high-high") or low values ("low-low") of two variables (dependent variable and lagged dependent variable) are spatially correlated given the existence of spillover effects. The regions with "high-low" and "low-high" values are "outliers" with negative spatial autocorrelation. In sum, this LISA Maps find clusters for the dependent variable and lagged dependent variable.

Upon analysing the Lisa Cluster Maps above (Figure 5), confirms what was seen with the Moran Scatterplots, or, in other words, only in the services with high values in the region around Greater Lisbon and low values in the Central region is there positive spatial autocorrelation. These figures also show some signs of positive spatial autocorrelation in all sectors, specifically with high values in the Greater Lisbon area and with low values in the Central Alentejo. On the other hand the industry presents signs of positive autocorrelation with high values in the Baixo Vouga in the Central region. In the second period (2000 to 2005) we can see differents situations what was expected, because the evolution of the Portuguese economy context was influenced by others factors, namely the national public and European supports.

**Figure 5: "LISA Cluster Map" of productivity for each of the economic sectors (cross-section analysis, 28 regions, 1995-1999)**

**a) Agriculture**             **b) Industry**

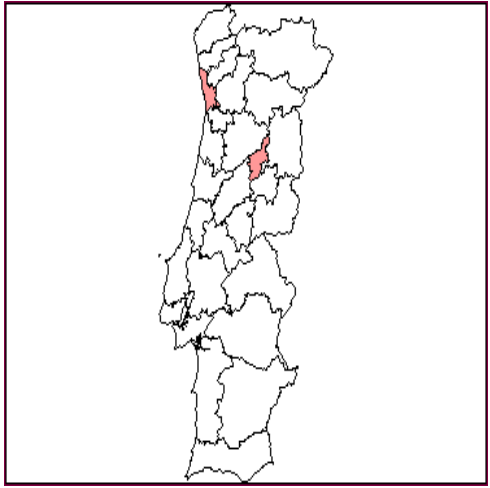 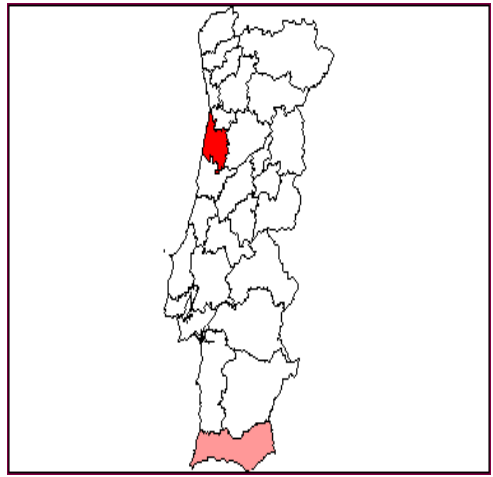

**c) Services**             **d) Total of sectors**

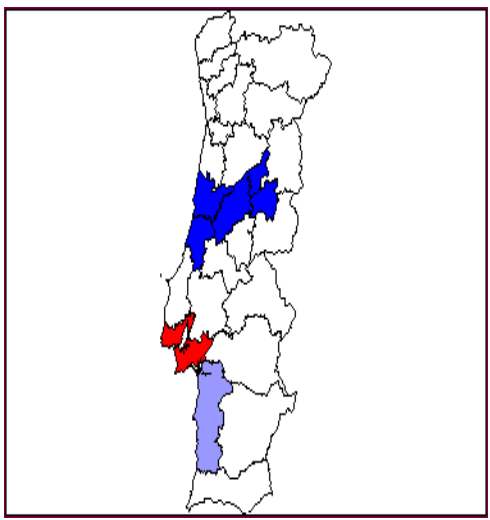 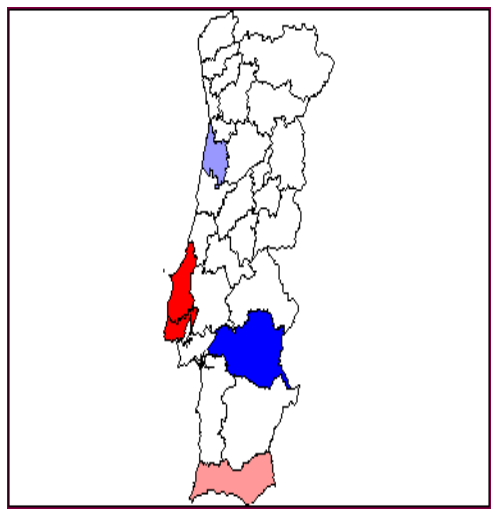

**Note:**

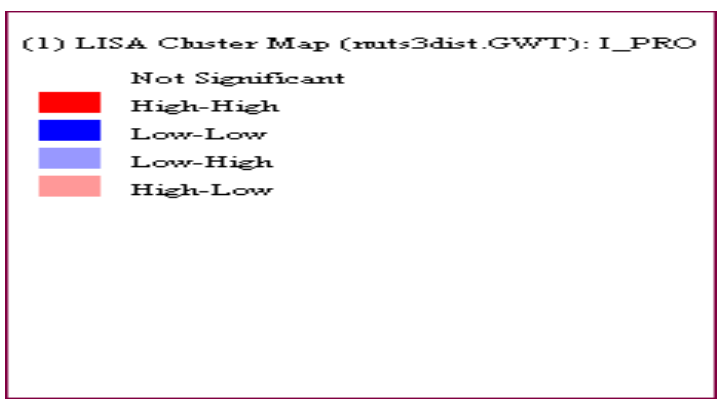

**Figure 6: "LISA Cluster Map" of productivity for each of the economic sectors (cross-section analysis, 28 regions, 2000-2005)**

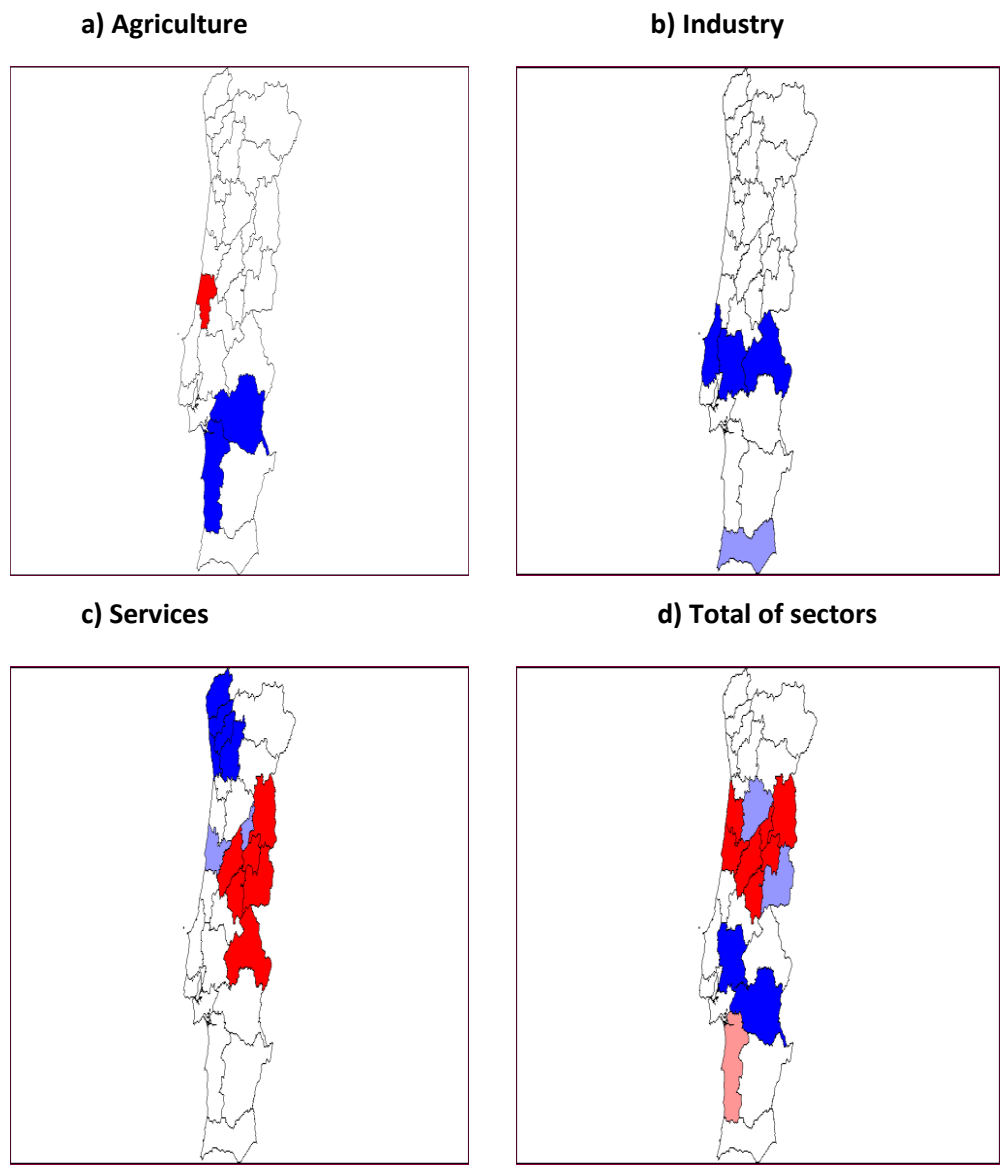

**Note:**

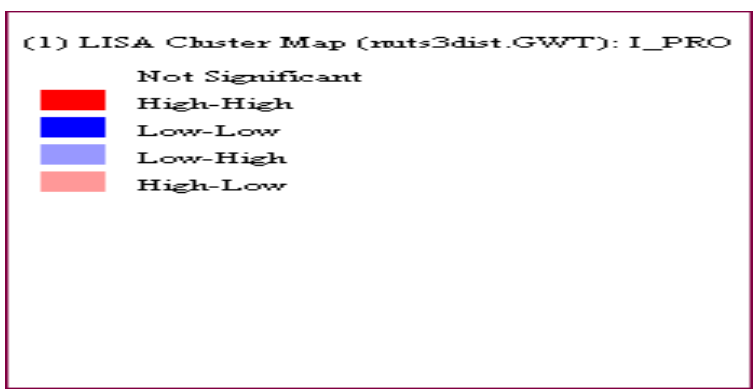

## 4. Econometric results

In the previous section we did an exploratory data analysis, using some tools of the GeoDa software. In this section we pretend analyse the relationship between the variables considered, with estimations, taking in count the economic theory and the econometric developments. So, with these estimations we can quantify, obtaining coefficients of regression, the influence of the independent variables (output growth and spatial effects) in the productivity growth.

The following presents empirical evidence of Verdoorn's relationship for each of the economic sectors in the Portuguese NUTs III from 1995 to 1999 and from 2000 to 2005, based on cross-section estimates. These cross- section estimates were carried out with the Least Squares (OLS) and the Maximum Likelihood (ML) methods.

This part of the study will examine the procedures of specification by Florax e al. (2006). So,we will firstly examine through OLS estimates and after the relevance of proceeding with estimate models with spatial lag and spatial error components with recourse to LM specification tests, like referred before.

The results concerning the OLS estimates in Veroorn's equation (equation (1), without spatial variables) with spatial specification tests are presented in Tables 1 and 2. In the columns concerning the test only values of statistical relevance are presented.

**Table 1: OLS cross-section estimates of Verdoorn's equation with spatial specification tests (1995-1999)**

Equation: $p_{it} = \alpha + \beta q_{it} + \mu_{it}$

| | Con. | Coef. | JB | BP | KB | M'I | $LM_l$ | $LMR_l$ | $LM_e$ | $LMR_e$ | $R^2$ | N.O. |
|---|---|---|---|---|---|---|---|---|---|---|---|---|
| **Agriculture** | 0.013* (3.042) | 0.854* (9.279) | 1.978 | 5.153* | 5.452* | 0.331* | 0.416 | 7.111* | 8.774* | 15.469* | 0.759 | 28 |
| **Industry** | -0.029* (-3.675) | 1.032* (9.250) | 3.380 | 2.511 | 1.532 | -0.037 | 1.122 | 2.317 | 0.109 | 1.304 | 0.758 | 28 |
| **Services** | 0.033* (3.971) | 0.169 (1.601) | 1.391 | 1.638 | 1.697 | 0.212* | 4.749* | 1.987 | 3.607* | 0.846 | 0.055 | 28 |
| **Total of sectors** | 0.002 (0.411) | 0.659* (8.874) | 1.585 | 5.174* | 4.027* | 0.030 | 0.008 | 0.087 | 0.069 | 0.149 | 0.742 | 28 |

**Note: JB, Jarque-Bera test to establish parameters; BP, Breusch-Pagan test for heteroskedasticity; KB, Koenker-Bassett test for heteroskedasticity: M'I, Moran's I statistics for spatial autocorrelation; $LM_l$, LM test for spatial lag component; $LMR_l$, robust LM test for spatial lag component; $LM_e$, LM test for spatial error component; $LMR_e$, robust LM test for spatial error component; $R^2$, coefficient of adjusted determination; N.O., number of observations; *, statistically significant for 5%**

**Table 2: OLS cross-section estimates of Verdoorn's equation with spatial specification tests (2000-2005)**

Equation: $p_{it} = \alpha + \beta q_{it} + \mu_{it}$

|  | Con. | Coef. | JB | BP | KB | M'I | LM$_l$ | LMR$_l$ | LM$_e$ | LMR$_e$ | R$^2$ | N.O. |
|---|---|---|---|---|---|---|---|---|---|---|---|---|
| **Agriculture** | -0.014 (-0.845) | 0.679* (2.263) | 1.201 | 0.300 | 0.505 | 0.108 | 0.771 | 0.030 | 0.940 | 0.198 | 0.132 | 28 |
| **Industry** | 0.015* (4.248) | 0.639* (6.572) | 3.238 | 2.703 | 1.393 | 0.236 | 8.742* | 4.366* | 4.444* | 0.068 | 0.610 | 28 |
| **Services** | -0.011* (-2.907) | 0.548* (3.841) | 2.728 | 9.579* | 10.452* | 0.227 | 5.976* | 1.998 | 4.102* | 0.124 | 0.338 | 28 |
| **Total of sectors** | 0.001** (0.079) | 0.513* (3.080) | 0.797 | 5.019* | 4.355* | 0.344 | 5.215* | 1.146 | 9.462* | 5.393* | 0.239 | 28 |

Note: JB, Jarque-Bera test to establish parameters; BP, Breusch-Pagan test for heteroskedasticity; KB, Koenker-Bassett test for heteroskedasticity: M'I, Moran's I statistics for spatial autocorrelation; LM$_l$, LM test for spatial lag component; LMR$_l$, robust LM test for spatial lag component; LM$_e$, LM test for spatial error component; LMR$_e$, robust LM test for spatial error component; R$^2$, coefficient of adjusted determination; N.O., number of observations; *, statistically significant for 5%

From the table 1 the existence of growing scaled income in agriculture and in the total of all sectors is confirmed, what is expected, taking into count the exploratory data analysis, namely for the agricultural sector. In this case we have a match between the two searchs, what is a signal of the robustness of the conclusions. Industry shows itself to be a sector with very strong growing scaled income, because, despite Verdoorn's coefficient being highly exaggerated, it is very close to unity. This is a confirmation of the Verdoorn Law and what Kaldor defended, in other words, the industry has increasing returns to scale and is the engine of the economy. As Kaldor predicts, services are subject to constant scaled income.. For this two sectors, the industry and the services, we confirm, too, what was concluded with the analysis of the data. We can see the existence of heteroscedasticity in agriculture and all sectors, given the values presented for these sectors by the BP and KB tests, but this is not a problem, since when we correct this statistic infraction with standardization the results are very close. As far as spatial correlation is concerned, Moran's value is only statistically significant in agriculture and services. For the services the Moran´s I value confirm what was seen with the exploratory analysis data. Following the procedure of Florax et al. (2006) the equation should be estimated with the spatial error component for agriculture and with the spatial lag component for services (although in this sector none of the robust LM tests have statistical significance), with the maximum likelihood method. As expected the values obtained for the second period are significantly different and in our point of view are better and more coherent with the economic theory. We see, also, which the results for the Verdoorn coefficients are very close of the conclusions obtained with the data analysis, what again show the robustness of the search. In this period, following the procedure of Florax et al. (2006) the equation should be estimated with the spatial error component for all sectors and with the spatial lag component for industry and services. On the other hand we see that the spillover effects have a negative influence in the productivity of the industry (Figure 6 and Table 4), what is a important problem which must be taking in count by the public policies, because this is a strategic economic sector for the economic growth.

The results for ML estimates with spatial effects for agriculture and services are presented in Tables 3 and 4.

**Table 3: Results for ML estimates for Verdoorn's equation with spatial effects (1995-1999)**

|  | Constant | Coefficient | Coefficient[(S)] | Breusch-Pagan | $R^2$ | N.Observations |
|---|---|---|---|---|---|---|
| **Agriculture** | 0.016* (1.961) | 0.988* (14.291) | 0.698* (4.665) | 4.246* | 0.852 | 28 |
| **Services** | 0.011 (0.945) | 0.134 (1.464) | 0.545* (2.755) | 3.050** | 0.269 | 28 |

Note: Coefficient[(S)], spatial coefficient for the spatial error model for agriculture and the spatial lag model for services; *, statistically significant to 5%; **, statistically significant to 10%.

**Table 4: Results for ML estimates for Verdoorn's equation with spatial effects (2000-2005)**

|  | Constant | Coefficient | Coefficient[(S)] | Breusch-Pagan | $R^2$ | N.Observations |
|---|---|---|---|---|---|---|
| **Industry** | 0.018* ( 5.535) | 0.682* (8.217) | -0.427* ( -2.272) | 4.103* | 0.714 | 28 |
| **Services** | -0.011* (-3.308) | 0.478* (3.895) | 0.533* (2.834) | 13.186* | 0.501 | 28 |
| **All the sectors** | -0.002 (-0.379) | 0.609* (4.328) | 0.616* (3.453) | 2.230 | 0.479 | 28 |

Note: Coefficient[(S)], spatial lag model for industry and services and the spatial coefficient for the spatial error model for the all sectors; *, statistically significant to 5%; **, statistically significant to 10%.

For the first period, it is only in agriculture that Verdoorn's coefficient improves with the consideration of spatial effects, since it goes from 0.854 to 0.988. On the other hand, the spillover effects have positive influence on the productivity growth of the agricultural sector and the services. In the second period, improve the Verdoorn´s coefficient of the industry and of the total sectors. In this way, despite the spatial spillover effects have negative influence in the productivity of the Portuguese industry, they improve the Verdoorn law. About the heteroscedasticity this is not a problem, since when we correct this statistic infraction with standardization the results are very close.

Is important to say, is not our case, but when we consider the presence of the spatial multiplier term in the spatial lag model the coefficients of the lag model are not directly comparable to estimates for the error model (Elhorts, 2010).

**5. Summary and concluding remarks**

Considering the analysis of the cross-section data previously carried out, it can be seen, for the first period, that productivity (product per worker) is subject to positive spatial autocorrelation in services (with high values in the Lisbon region and low values in the Central region) and in all sectors (with high values in the Lisbon region and low values in the Central Alentejo) and also in industry (although this sector has little significance, since high values are only found in the NUT III Baixo Vouga of the Central Region). Therefore, the Lisbon region clearly has a great influence in the development of the economy with services. On the other hand, what Kaldor defended is confirmed or, in other words Verdoorn's relationship is stronger in industry, since this is a sector where growing scaled income is most expressive.

As far as cross-section estimates are concerned, it can be seen, for the first period also, that sector by sector the growing scaled income is much stronger in industry and weaker or non-existent in the other sectors, just as proposed by Kaldor. With reference to spatial autocorrelation, Moran's I value is only statistically significant in agriculture and services. Following the procedures of Florax et al. (2006) the equation is estimated with the spatial error component for agriculture and the spatial lag component for services, it can be seen that it is only in agriculture that Verdoorn's coefficient improves with the consideration of spatial effects.

For the second period the data and the results are different, what is waited, because the context in Portugal is distinct and in our point of view the indicators are better. The data and the results are more coherent with the economic theory, namely about the Verdoorn coefficient. Verdoorn in his work found values for the his coefficient about 0,45. The values for this second period are close of these results principally in the industry. On the other perspective, the results for the spillover effects are more well distributed for the several economic sectors, despite the case of the industry, what is a good sign in the sense of the Portuguese economy modernization.

In a policy decisions way is important continue the structural process of modernization of the Portuguese economy and the national public and European supports for the economic sectors improvement are much important. Namely, for the industry that is the strategic economic sector for the national economic growth, considering what Kaldor said and because is a sector that produce tradable products. The policies must be directed for manufacturing industrial sectors with high technology with great increasing returns to scale. We must built policies that impulse the appearance of new industries and try to improve what we have a lot and have low increasing returns to scale industries like the textile and shoes sectors. Maybe a improvement in the employers and employees formation will be a good sign to the modernization of Portuguese industry, because we saw that when there spatial spillover effects in this sector the Verdoorn law improve, but we found low influence of these effects in the industry. This is a sign that the spillover do not spread in the industry of the Portuguese regions and this is a sign that the firms are not able to take the benefits of the neighbours.

For the future is important try analysing the Verdoorn law and the existence of spatial spillover effects for the several manufacturing industry of the Portuguese regions to see what happen at the different industrial sectors. On other way will be interesting considering others variables in the model, like the human capital, their formation improve and the several policies, to try find others influences in the way of the modernization the Portuguese economy.